\documentclass[a4paper,12pt]{article}
\usepackage[dvips]{graphicx}
\usepackage{color}
\usepackage{epsf}
\usepackage{amssymb}
\usepackage{amsmath}

\usepackage{lscape}
\usepackage{tabularx}
\usepackage{booktabs}

\makeatletter 
\def\section{\@startsection {section}{1}{\z@}{-3.5ex plus -1ex minus -.2ex}{2.3 ex plus .2ex}{\Large\bf}}
\makeatother

\def\argmax{\mathop{\mathrm{argmax}}}

\newtheorem{example}{Example}

\begin{document}

\title{Fighting against uncertainty:\\ 
  An essential issue in bioinformatics\footnote{This manuscript was accepted in {\bf Briefings in Bioinformatics} for publication}}
\author{
  Michiaki Hamada\,$^{1,2}$\footnote{To whom correspondence should be addressed.
    Tel.: +81-3-5281-5271; Fax: +81-3-5281-5331; E-mail: mhamada@k.u-tokyo.ac.jp}\\
  $^{1}$the University of Tokyo, 
  $^{2}$CBRC/AIST}
\pagestyle{plain}
\date{\today}
\maketitle
%
\begin{abstract}
  Many bioinformatics problems, such as sequence alignment, gene prediction, 
  phylogenetic tree estimation and RNA secondary structure prediction, 
  are often affected by the ``uncertainty'' of a solution;
  that is, 
  the probability of the solution is extremely small. 
  This situation arises for estimation problems on high-dimensional discrete spaces in which
  the number of possible discrete solutions is immense. 
  In the analysis of biological data or the development of prediction algorithms, 
  this uncertainty should be handled carefully and appropriately. 
  In this review, I will explain several methods to combat this uncertainty,  
  presenting { a number of} examples in bioinformatics.
{
  The methods include (i) avoiding point estimation,
  (ii) maximum expected accuracy (MEA) estimations,
  and (iii) several strategies to design a pipeline involving several prediction methods.}
  I believe that the basic concepts and ideas described in this review will 
  be generally useful for estimation problems in {various} areas of 
  bioinformatics.
\end{abstract}
%
%
\section{Introduction}\label{sec:intro}
%
\subsection{Uncertainty of solutions}\label{sec:uncertainty}
\begin{figure}[th]
  \centerline{
    \includegraphics[width=0.75\textwidth]{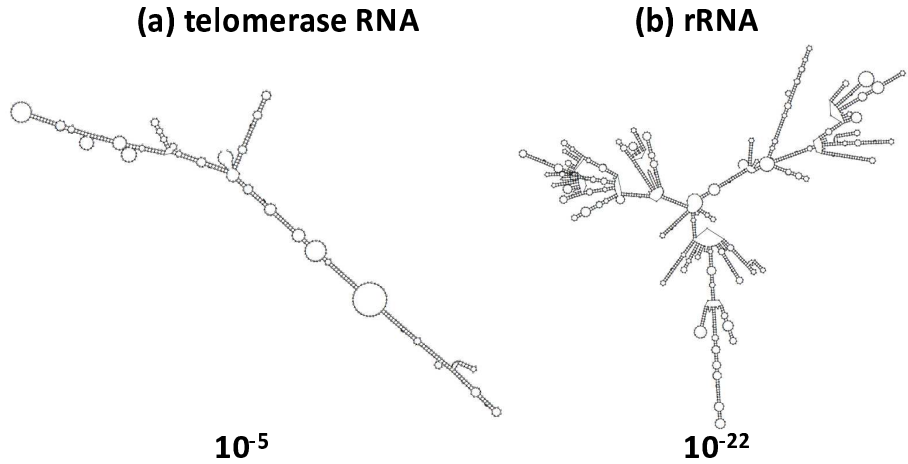}
  }\caption{\label{fig:mfe_sec}\small
    The probabilities  of forming 
    the minimum free energy (MFE) 
    structure, computed by using the Boltzmann distribution (see Eq~\ref{eq:BD}),
    for (a) telomerase RNA and (b) ribosomal RNA.
    Note that both probabilities are the {\em highest} ones 
    among all possible secondary structures for each RNA,
    because the structures have the {minimum} free energy 
    of all possible structures.
  }
\end{figure}

%
In estimation problems that appear in bioinformatics,
such as 
sequence alignment \cite{durbin1998biological}, 
gene prediction \cite{pmid22510764},
RNA secondary structure prediction \cite{pmid22736001,Aigner},
RNA-RNA interaction prediction \cite{pmid20823308},
and 
phylogenetic tree estimation \cite{Yang2006}, 
it is typical that the probability of any solution 
is extremely small even if that solution has the highest probability.
For example, the probabilities of the minimum free energy (MFE) structures 
of a telomerase RNA and a ribosomal RNA are 
$8\times 10^{-6}$ and $4 \times 10^{-23}$, respectively\footnote{%
This example shows that one needs to be careful of 
predicted secondary structures for
long RNA sequences using simple minimum free minimization. In practical, a 
method recently proposed by \cite{pmid23511969} (or \cite{pmid14734309}) would partially mitigate this issue.} 
(Figure~\ref{fig:mfe_sec}).
Here the probabilities were computed by using the Boltzmann distribution 
(also called the Gibbs distribution) from statistical mechanics \cite{pmid1695107}:
%
\begin{align}
  p(\theta|x)=\frac{1}{Z(x)} \exp \left( \frac{-E(\theta,x)}{RT} \right)\label{eq:BD}
\end{align}
%
where $\theta$ is a specific secondary structure (e.g., the MFE structure) of 
an RNA sequence $x$, $E(\theta,x)$ is the free energy 
of secondary structure $\theta$, 
$T$ is the temperature, 
$R$ is the ideal gas constant (8.31 J/mol K) and $Z(x)$ is the normalizing
constant (partition function) that ensures $\sum_\theta p(\theta|x)=1$.
The situation is much worse in the problem of RNA deleterious mutation prediction \cite{pmid21422070,pmid16638137,pmid18688270},
in which a specific RNA secondary structure is sought from among
all possible secondary structures with up to $k$ mutations and the partition function 
thus becomes much larger (cf. see Figure 4 in Barash and Churkin \cite{pmid21422070}).

The fact that the probability of a solution can be extremely small is known as the {\em uncertainty} of solutions, 
which leads to a critical issue in developing prediction algorithms or analyzing biological data 
in the field of 
bioinformatics~\cite{pmid18073381,pmid23160142,pmid18218900,pmid20447933,pmid21409513}.
In one \textit{Science} paper \cite{pmid18218900}, for instance,
the authors argued that 
the uncertainty of multiple sequence alignment greatly influences 
phylogenetic topology estimations:
phylogenetic topologies estimated from multiple alignments predicted by 
5 widely used aligners are different from one another.
There is no doubt that this raises serious issues for reaching
biological conclusions.
%

\subsection{Where does the uncertainty come from?}\label{sec:why}
%
The uncertainty described in the previous section 
arises for estimation problems 
on high-dimensional discrete spaces, in which the number of possible solutions 
is immense.
For instance, the following results for the asymptotic number of solutions in particular problems are known:
%
\begin{itemize}
\item The number of possible secondary structures, $S(n)$, of an RNA sequence with 
  length $n$ is estimated to be
  \[
  S(n)\sim \sqrt{\frac{15+7\sqrt{5}}{8\pi}}n^{-1.5}\left(\frac{3+\sqrt{5}}{2}\right)^n
  \]
  where $f(n)\sim g(n)$ means that $\lim_{n\to \infty} f(n)/g(n)=1$\footnote{%
    The number of secondary structures of a specific RNA sequence 
    will generally be smaller than this value 
    because, in the equation, it is assumed that any pair of nucleotides can 
    form a base-pair. 
  }.
  See Theorem 13.2 in \cite{citeulike:3517977} for the proof.
\item 
  Let $f(n,m)$ be the number of alignments of one sequence of $n$ letters with
  another of $m$ letters. Then $f(n,n)$ is estimated to be
  \[
  f(n,n)\sim(1+\sqrt{2})^{2n+1}n^{-1/2}.
  \]
  See Theorem 9.1 in \cite{citeulike:3517977} for the proof.
\item 
  The number of topologies of a phylogenetic tree (un-rooted binary tree) 
  with $n(>3)$ leaves 
  (formally called ``operational taxonomy units'') is 
  $\prod_{j=3}^n (2j-5)=1\cdot 3 \cdot 5 \cdot \cdots \cdot (2n-5)$.
  See Proposition 14.1 in \cite{citeulike:3517977} for the proof.
\end{itemize}

In the equations above, 
the number of solutions increases exponentially with $n$ 
so, for example, the number of alignments of two sequences with length 1000 
is $ f(1000,1000) \sim 10^{767.4\ldots}$, which is immense because
it has been {estimated} that the number of particles in the universe is about $10^{80}$.
Because the sum of the probabilities of all solutions must be equal to 1,
the probability of any particular solution tends to become extremely small because of the large number of potential solutions.
In addition, there often exist many very-similar solutions: for a given RNA secondary structure $s$, secondary structures produced by unpairing a base-pair (in $s$) are similar to the original secondary structure.

\subsection{Purpose of this review}\label{sec:purpose}
%
In this review, I aim to introduce several methods to handle 
this uncertainty appropriately in the field of bioinformatics, 
presenting a number of actual studies. Most examples are adopted from either RNA informatics 
or sequence alignments.
{
  The methods include (i) avoiding point estimation (e.g., prediction of suboptimal solutions),
  (ii) maximum expected accuracy (MEA) estimations, and 
  (iii) several strategies to design a pipeline involving several prediction methods.}

Throughout this article, the following notation and definitions are used.
%
\begin{itemize}
\item $Y$ denotes a {\em predictive (solution) space}, including all the possible 
  solutions that are discrete. 
  For example, $Y$ is the set of possible RNA secondary structures of an RNA sequence $x$.
\item 
  $p(\theta|D)$ denotes a posterior probability distribution 
  on a predictive space $Y$, given data $D$. 
  For example, $p(\theta|x)$ is a probability distribution of secondary structures 
  of an RNA sequence $D=\{x\}$ given by Eq.~(\ref{eq:BD}).
  In this study, I will just assume the existence of this probability distribution 
  on the predictive space in most cases.
\item
  Predicting one solution from the predictive space $Y$ is called a {\em point estimation}.
  The probability of the solution found by point estimation is sometimes extremely low 
  as described in the previous section.
\end{itemize}

This paper is organized as follows.
In Section~\ref{sec:giveup}, 
I describe methods to avoid point estimation by sampling several 
sub-optimal solutions or 
visualizing the distribution of solutions.
In Section~\ref{sec:one_solution}, I introduce methods 
to predict one solution, taking the uncertainty into account, 
because there are several situations 
in which only one solution is required.
In Section~\ref{sec:pipeline}, I give some strategies for
developing complex pipelines or algorithms in which several prediction methods are 
involved.
In Section~\ref{sec:discussion} these approaches are discussed 
and the direction of future research in areas related to this study is considered.

\section{Avoiding point estimations}\label{sec:giveup}
%
\subsection{Prediction of suboptimal solutions}\label{sec:suboptimal}
%
One possible method to handle the {uncertainty} described in Section~\ref{sec:intro} 
is to predict not only the one {\em optimal} solution\footnote{%
  There might be {\em several} optimal solutions whose score 
  is exactly the same as the optimal score.
  In this case, prediction of {\em all} optimal solutions also raises an issue. 
} but also several 
{\em sub-optimal} solutions.
{%
  Zuker \cite{pmid2468181}, in a study of RNA secondary structure prediction,
  was the first to introduce a method
  to predict suboptimal solutions, implemented in Mfold \cite{pmid12824337} (See Supplementary Section~\ref{sec:mfold} for the details).
  %
}
%
\subsubsection{Stochastic sampling with dynamic programming}\label{sec:ss}
%
Stochastic sampling (SS) in combination with a dynamic programming (DP) technique enables the efficient sampling of solutions from a distribution 
$p(\theta|D)$ of solutions.
For both RNA secondary structure predictions and pairwise alignments, 
stochastic sampling is realized by stochastically 
conducting a traceback procedure 
in a DP algorithm \cite{durbin1998biological,pmid10404626}. 
For example, RNAsubopt \cite{pmid10070264} predicts the complete set of
suboptimal secondary structures within a given energy range 
(say, 1 kcal/mol) from 
the minimum free energy.

\subsubsection{MCMC and Gibbs sampling}

For a probability distribution whose probabilistic structure 
is more complicated,
a Markov-chain Monte-Carlo (MCMC) algorithm \cite{citeulike:4879195}, such as Gibbs sampling, is
a widely applicable method to sample from the posterior distribution, although
in most applications MCMC is used to obtain the  optimized solution.

Meyer {\it et al.}~\cite{pmid17696604} applied 
an MCMC method to optimize the joint probability 
distribution
\begin{align}
  P(S,A,T|D)=\frac{1}{Z} P(D|S,A,T)P(S,A,T)
\end{align}
where $D$ stands for the data (i.e., unaligned RNA sequences), 
$Z=P(D)$ is a normalizing constant (partition function),
$S$ is a consensus (or common) RNA secondary structure that might contain
pseudoknots, $A$ is a multiple sequence alignment, and $T$ is an
evolutionary tree relating the sequences. 
It would be difficult not only to optimize this posterior probability using
dynamic programming techniques
but also to apply stochastic sampling with a dynamic programming 
described in the previous section, because the probability structures 
are complex joint distributions of three states. 

Metzler and {Nebel}~\cite{pmid17589847} and Bon {\it et al.}~\cite{pmid23248008}
proposed a probabilistic model for RNA secondary structures 
with pseudoknots and presented an MCMC Method 
for sampling RNA structures according to their posterior distribution 
for a given sequence.
In contrast to conventional RNA secondary structure prediction 
(which does not consider pseudoknots), 
RNA secondary structure prediction with pseudoknots 
entails a larger computational cost (see \cite{pmid21548808}) and
the application of stochastic sampling is not realistic.
%
Doose and Metzler~\cite{pmid22796961} presented a method (PhyloQfold) 
that takes advantage
of the evolutionary history of a group of aligned RNA sequences for
sampling consensus secondary structures, including pseudoknots,
according to their approximate posterior probability.
%
Wei {\it et al.}~\cite{pmid21788211}
presented a new global structural alignment algorithm, RNAG, to predict consensus
secondary structures for unaligned sequences, utilizing a blocked
Gibbs sampling algorithm.

For phylogenetic tree estimations,
MrBayes \cite{pmid11524383} is a program for Bayesian phylogenetic 
analysis, which uses MCMC techniques 
to sample from the posterior probability distribution of phylogenetic trees for a given
multiple sequence alignment.
More recently, the program BigFoot \cite{pmid19715598} employed an MCMC method for finding
joint distributions of phylogenetic trees and multiple sequence alignments (MSA).

As shown by the above examples, MCMC and Gibbs sampling seem
useful methods to sample suboptimal solutions when the structure of target probability distribution 
is complex.
%
\subsubsection{Non-stochastic approaches}

\begin{figure}[t]
  \centerline{
    \includegraphics[width=0.6\textwidth]{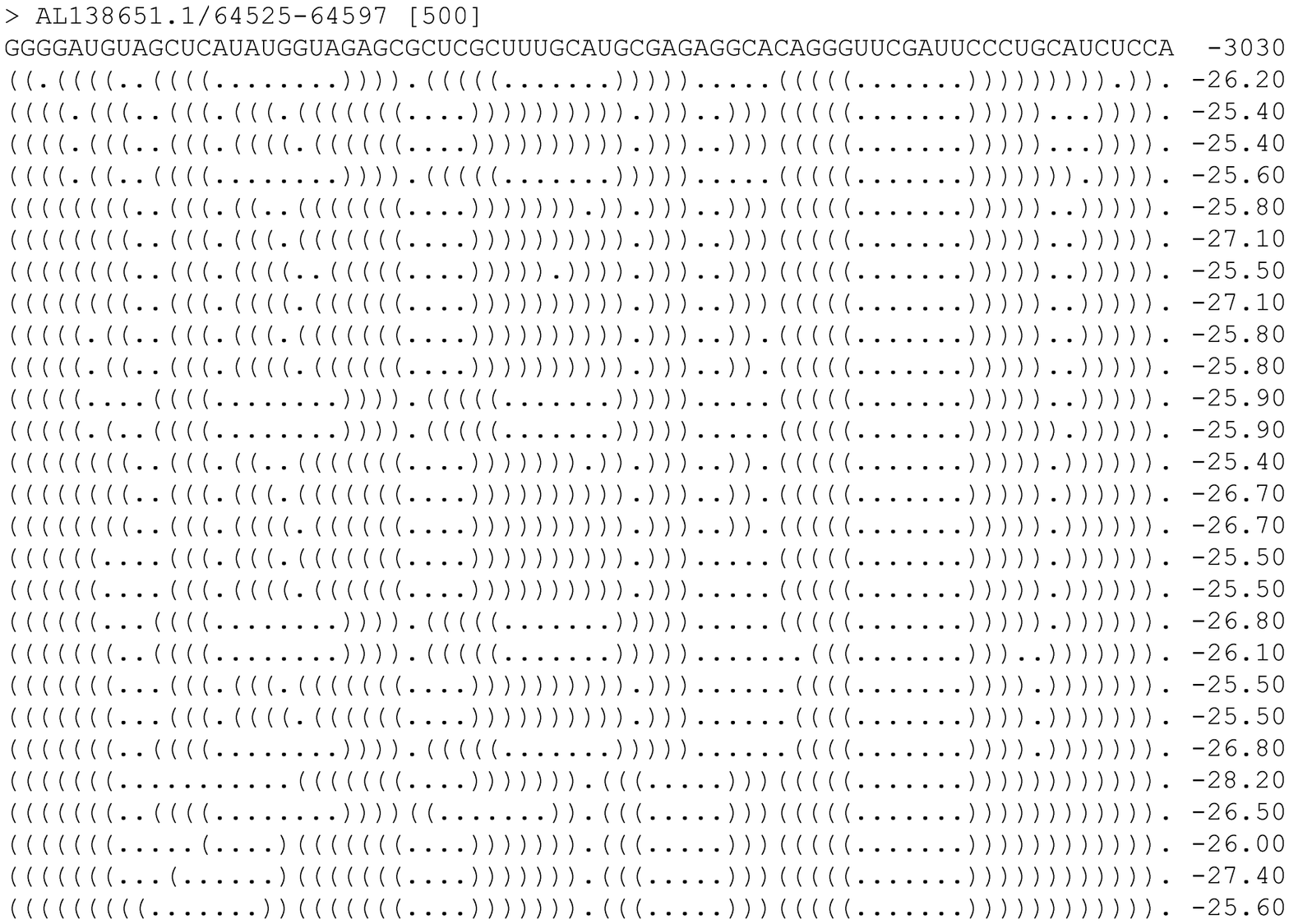}
  }\caption{\label{fig:rnasubopt}\small
    Suboptimal secondary structures of a tRNA sequence  predicted by 
    RNAsubopt \cite{pmid10070264}, enumerating all possible suboptimal RNA secondary structures within
    a given energy range from the minimum free energy (MFE). 
  }
\end{figure}

\begin{figure}[t]
  \centerline{
    \includegraphics[width=0.9\textwidth]{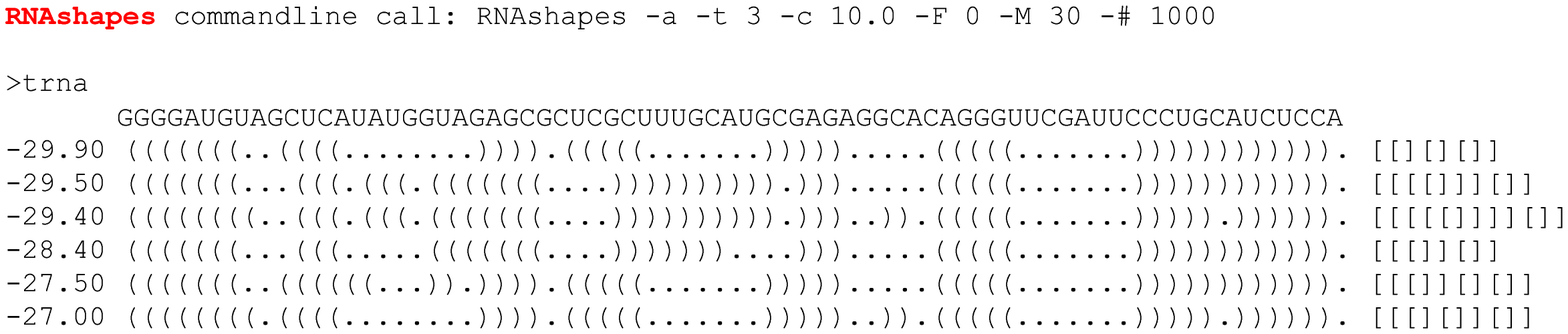}
  }
  \caption{\label{fig:rnashapes_trna}\small
    RNAshapes (http://bibiserv.techfak.uni-bielefeld.de/rnashapes/) 
    \cite{pmid16357029} results for secondary structures of a tRNA sequence. 
    The rightmost structures of each secondary structure with `[' and `]' are
    abstract shapes, which represent coarse-grain secondary structures. RNAshapes predicts
    only one optimal secondary structure for each abstract shape, reducing the number of
    predictions.
  }
\end{figure}

Typically, the approach { described in the previous sections} 
returns a huge number of secondary structures 
in which there are many structures that are similar to each other 
(cf. Figure~\ref{fig:rnasubopt}).
To overcome this, a method to predict representative 
suboptimal secondary structures, such as locally optimal secondary structures or alignments, 
has been proposed \cite{pmid2468181,pmid15725735,pmid23057822} {(cf. Supplementary Section~\ref{sec:mfold})}.
Moreover, for RNA secondary structure predictions, 
RNAshapes \cite{pmid16357029} implements an algorithm to predict secondary structures
for every {\em abstract shape}, { which is realized by using an algebraic dynamic programming (ADP) technique \cite{Giegerich02algebraicdynamic}}. 
This { reduces} the number of predictions
(Figure~\ref{fig:rnashapes_trna}). 
Note that a probabilistic version of RNAshapes has also been proposed, which
computes the accumulated probabilities of all structures that share a shape \cite{pmid16480488}.

%

\subsubsection{Prediction of representative solutions after clustering}
\begin{figure}[t]
  \centerline{
    \includegraphics[width=0.99999\textwidth]{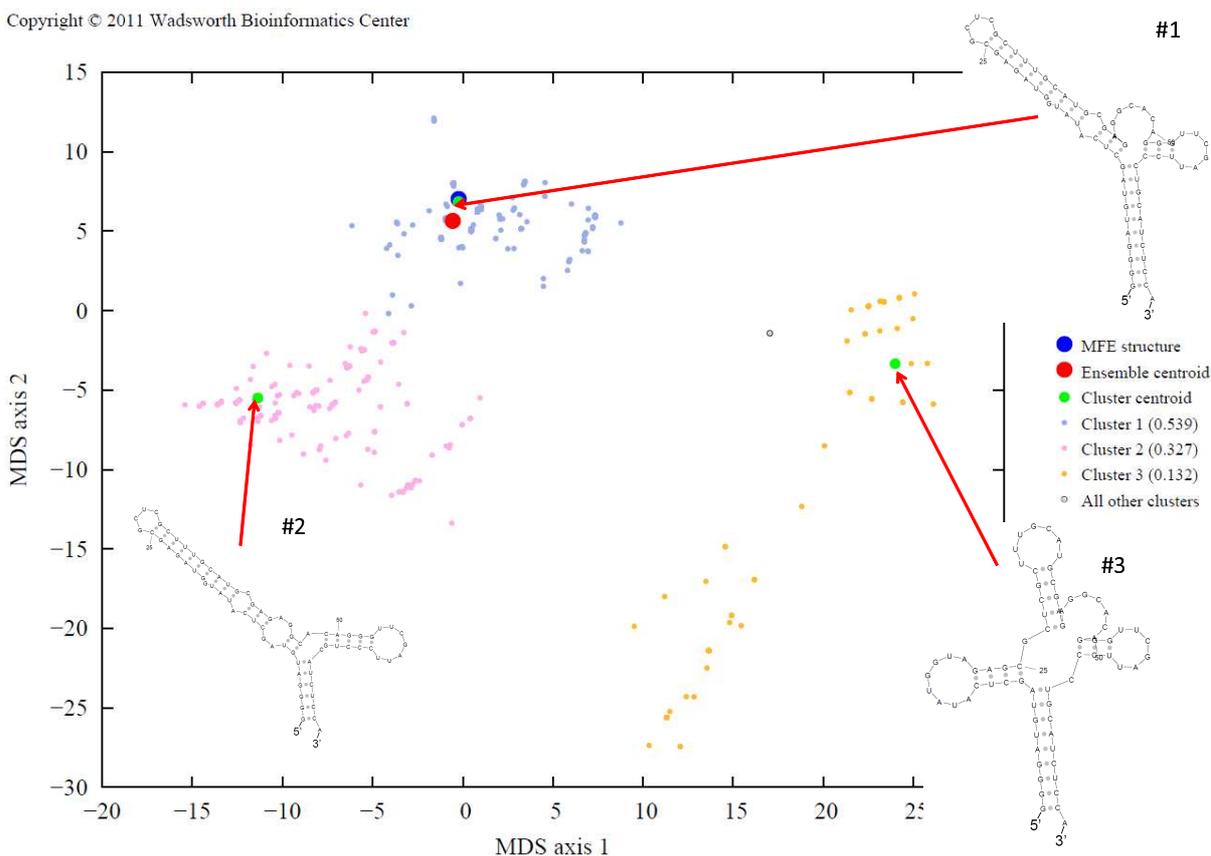}
  }
  \caption{\label{fig:MDS_sfold}\small
    Clustering result from the Sfold Web Server 
    (http://sfold.wadsworth.org/) \cite{pmid16043502}.
    A tRNA sequence was used to draw this figure.
    The space of secondary structures is embedded into two dimensional space using 
    Multidimensional distance scaling \cite{kruskal&wish:78} (the center plot) 
    on the base-pair distance of RNA secondary structures.
    Four secondary structures are shown, corresponding to the centroid structures
    for the 1st, 2nd and 3rd clusters, respectively.
  }
\end{figure}

%
In this approach,
first stochastic sampling (Section~\ref{sec:ss}) is performed,
then the suboptimal solutions are clustered, and finally 
a solution for each cluster is predicted.
In this way, it is expected that a diverse variety of solutions 
(i.e., solutions are not similar to each other) will be obtained.
For RNA secondary structure predictions,
this approach is implemented in Sfold \cite{pmid16043502} and CentroidFold \cite{pmid19435882}, 
Sfold provides a Web-interface for this approach
(http://sfold.wadsworth.org/cgi-bin/index.pl). See Figure~\ref{fig:MDS_sfold} for
an example of the output of the Sfold Web Server for a typical tRNA sequence.

\subsection{Visualizing distributions of solutions}\label{sec:vis}

Visualization of the distribution of solutions 
would include much richer information than a single solution.
Yet, visualizing solutions is not trivial 
because the solution space is
generally high-dimensional (cf. Section~\ref{sec:why}).

\subsubsection{Visualizing distributions of solutions with sampling}\label{sec:vis_ent}
%
In many cases, a solution lies in a high-dimensional discrete space and 
the number of possible solutions is immense;
it is therefore difficult to visualize solutions directly.
To address this,
multidimensional distance scaling (MDS) \cite{kruskal&wish:78} or 
principal component analysis (PCA) is utilized  
in combination with the sampling of solutions by stochastic sampling or MCMC.

For RNA secondary structure predictions, Sfold produces this kind of visualization 
of a distribution 
of RNA secondary structures in terms of base-pair distance\footnote{The base-pair distance is equal 
  to the number of base-pairs that differ between two secondary structures \cite{pmid16043502}.} (Figure~\ref{fig:MDS_sfold}). 

For phylogenetic tree estimations,
Amenta and Klingner \cite{1173150} and
Hillis {\it et al.} \cite{pmid16012112} proposed a method to visualize
phylogenetic trees 
by utilizing MDS 
based on a distance between two phylogenetic trees,
called the Robinson-Foulds (RF) distance \cite{Robinson1981131} (the sum of the internal edges in 
disagreement between the two trees).
More recently, Huang and Li \cite{pmid23294272}
have developed some software MASTtreedist that employs MDS with the maximum agreement subtree (MAST) distance, which is the number of leaves in common 
for the maximum subtree between the two trees \cite{Bryant97buildingtrees}. This 
leads to better clustering results than 
using the RF distance.

\subsubsection{Visualizing distributions with one or two reference solutions}\label{sec:dist_sol}

Instead of visualizing the distribution of solutions directly (Section{\ref{sec:vis_ent}), 
  it has been proposed to visualize the distribution with one or two reference solutions.
  Let us first consider a reference case, in which one reference solution $s_0$ 
  and a distance, $d(\cdot,\cdot)$,
  between two solutions  are given.
  For each integer value $k$, 
  $p(k)$ is equal to the sum of probabilities for solutions whose distance from $s_0$ is equal to $k$:
  \begin{align*}
    p(k)=\sum_{s'\in \{s| d(s_0,s)=k \}} p(s'|D)
  \end{align*}
  where $p(\cdot|D)$ is a probability distribution over solutions.
  Then, the distribution $p(k)$ ($k=1,2,3,\cdots$) can be easily visualized.
  One advantage of this visualization is that, 
  for several problems such as alignments and
  RNA secondary structure predictions, the distribution 
  can be computed exactly by employing dynamic programming 
  techniques
  (Newberg and Lawrence \cite{pmid19119992})\footnote{The approach described in Section~\ref{sec:vis_ent} 
    requires candidate solutions to be visualized. These candidate solutions are usually given by sampling techniques because
    it is infeasible to visualize the complete distribution.}.
  The programs RNAbor \cite{pmid17526527,pmid17573364} and RNAborMEA \cite{pmid22537010} implement this 
  approach.
  Figure~\ref{fig:rnabor} shows an example of RNAbor results for a 101 nucleotide SAM riboswitch.
  It should be noted that points with a large base-pair distance from the 
  reference structure
  may include a lot of structures which 
  might be very different from each other.

  This approach can be extended to two (or more) solutions. 
  If two reference solutions $s_0$ and $s_1$ are given,
  the distribution $p(k,l)$ ($k,l=1,2,\cdots$) is defined by
  \begin{align*}
    p(k,l)=\sum_{s'\in \{s| d(s_0,s)=k \wedge d(s_1,s)=k \}} p(s'|D).
  \end{align*}
  In this case, the distribution becomes more specific and includes richer information than
  that with one reference case. The entire distribution $p(k,l)$ with two reference solutions 
  can be efficiently
  computed by using dynamic programming for RNA secondary structures and alignments \cite{DBLP:conf/gcb/LorenzFH09}.

  Recently, a more efficient method for visualization of both one and two reference cases has been
  proposed. This method 
  uses a discrete Fourier transform (DFT) and parallel computing 
  (Mori {\it et al.}, manuscript in preparation).
  %
  \begin{figure}[t]
    \centerline{
      \includegraphics[width=0.85\textwidth]{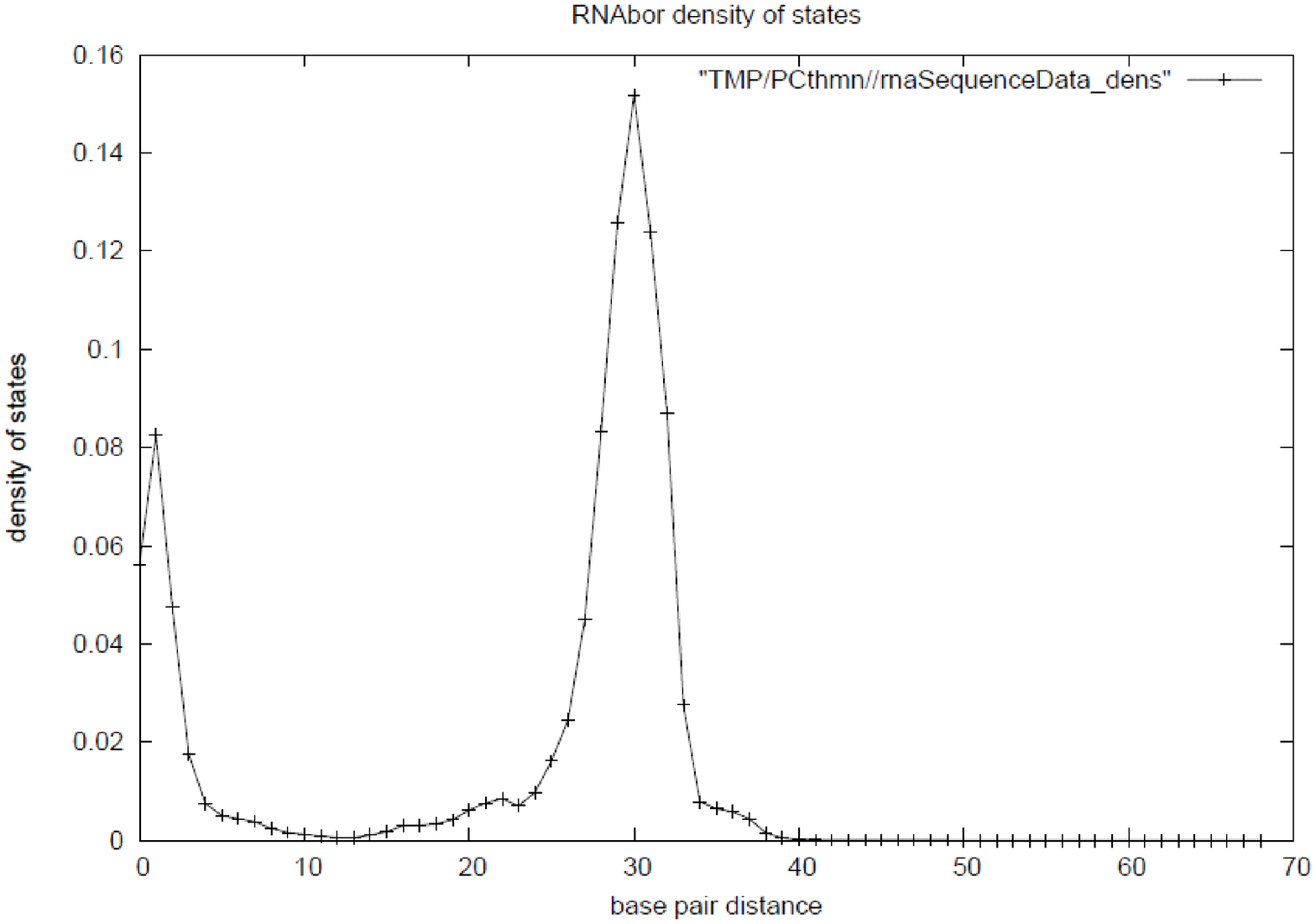}
    }\caption{\label{fig:rnabor}\small
      The result of RNAbor for a
      101 nucleotide SAM riboswitch sequence from the Rfam database \cite{pmid23125362}, 
      with EMBL accession code AP004597.1/118941-119041: 
      {\scriptsize UACUUAUCAAGAGAGGUGGAGGGACUGGCCCGCUGAAACCUCAGCAACAGAACGCAUCUGUCUGUGCUAAAUCCUGCAAGCAAUAGCUUGAAAGAUAAGUU}.
      The secondary structure corresponding to the origin is the reference structure in the Rfam database.
      The horizontal axis indicates base-pair distance \cite{pmid16043502}, and
      the vertical axis indicates probability density of RNA secondary structures whose base-pair distance is
      equal to the value on the horizontal axis.
    }
  \end{figure}

  \subsection{Visualizing marginal probabilities}\label{sec:marg}
  %
  Marginal probabilities are the sums of probabilities of solutions that 
  satisfy a specific condition, and are formally represented by
  \begin{align}
    p_\mathcal{C} = \sum_{\theta\in \mathcal{C}} p(\theta|D)
  \end{align}
  where $\mathcal{C}$ is a subset of the predictive space $Y$ satisfying a condition of interest.
  Marginal probabilities with respect to an appropriate condition
  are often much larger (typically, near to 1) than the probability 
  of a solution itself
  (which is, typically, less than $10^{-5}$; see Figure~\ref{fig:mfe_sec}), 
  and visualization of marginalized probabilities includes rich information, as shown in the following examples.
  %
  \begin{example}[Base-pairing probabilities (BPPs)]\label{eg:bpp}
    A base-pairing probability (BPP), $p_{ij}$, of an RNA sequence $x$ is the marginal probability 
    that $x_i$  and $x_j$ (the $i$-th and $j$-th nucleotides of $x$) form a base-pair:
    \begin{align}
      p_{ij} = \sum_{\theta\in \mathcal{C}(i,j)} p(\theta|x)\label{eq:bpp}
    \end{align}
    where
    $\mathcal{C}(i,j)$ is a set of secondary structures whose $i$-th position forms
    a base-pair with the $j$-th position, and
    $p(\theta|x)$ is a probability distribution of secondary structures of $x$, for example, 
    given by 
    the McCaskill model \cite{pmid1695107} (cf. Eq.~(\ref{eq:BD})).
  \end{example}

  The set of all base-pairing probabilities for all pairs of nucleotides 
  in a given RNA sequence
  is called a ``base-pairing probability matrix'' (BPPM), and is represented as a triangular
  matrix $P=\{p_{ij}\}_{1\le i < j \le |x|}$ where $p_{ij}\in[0,1]$ is defined in Eq.~(\ref{eq:bpp}).
  The BPPM of typical models for RNA secondary structures can be computed in a polynomial order time by
  employing a dynamic programming algorithm (e.g., see \cite{pmid1695107}).

  In Figure~\ref{fig:bppm_apt}, the base-pairing probability matrix for an RNA aptamer
  is shown. 
  %
  Notice that, although the base-pairing probability matrix does not specify a 
  secondary structure, it includes richer information about structures 
  than can be obtained from a single RNA secondary structure. See Section~\ref{sec:useful_marg_anal} for the details.

  Recently, other visualizations of base-pairing probabilities have been proposed 
  (RNABow \cite{pmid23407410}; http://rna.williams.edu/rnabows/single.html). See {Supplementary} Figure~\ref{fig:rnabow} for an example of RNABow output.

  \begin{figure}[t]
    \centerline{
      \includegraphics[width=0.5\textwidth]{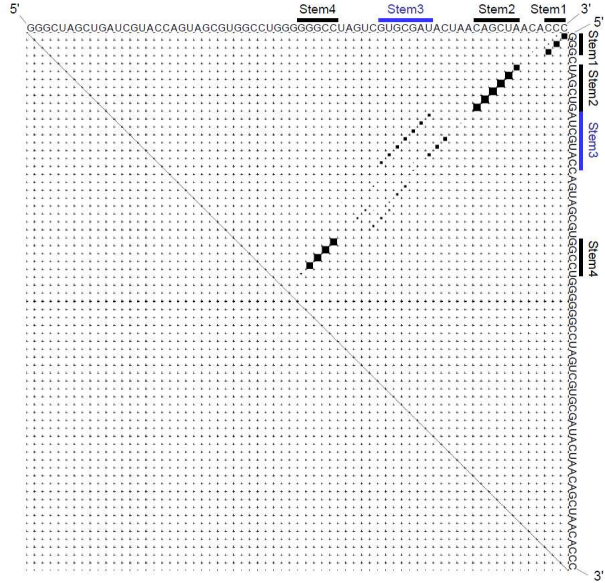}
    }
    \caption{\label{fig:bppm_apt}\small
      A base-pairing probability matrix for an RNA aptamer \cite{pmid21524680}.
      See Figure~\ref{fig:apt_bppm} for the detailed implication of it.
      by biochemical experiments. 
    }
  \end{figure}
  \begin{example}[Aligned-pairing probabilities (APPs)]\label{eg:app}
    An aligned-pairing probability (APP), $p_{ik}$, of two sequences $x$ and $x'$ 
    gives the probability that $x_i$ and $x'_k$ align: 
    \begin{align}
      p_{ik} = \sum_{\theta\in \mathcal{C}(i,k)} p(\theta|x,x')\label{eq:app}
    \end{align}
    where
    $\mathcal{C}(i,k)$ is the set of pairwise alignments whose $i$-th base 
    (nucleotide or amino acid) in $x$ aligns with
    the $k$-th base in $x'$, and
    $p(\theta|x,x')$ is a probability distribution for pairwise alignments between $x$ and $x'$, 
    such as the Miyazawa model \cite{pmid8771180} or pair hidden Markov models (pHMMs) \cite{durbin1998biological}.
  \end{example}
  %

  The entire set of probabilities for every possible pair of base alignments 
  between two sequences $x$ and $x'$ is called an
  ``aligned-pairing probability matrix'' (APPM), 
  It is represented
  by a matrix $P=\{p_{ik}\}_{1\le i \le |x|, 1\le |x'|\le k}$ 
  where $p_{ik} \in [0,1]$ is defined in Eq.~(\ref{eq:app}).
  The APPM of typical models for pairwise alignments can be computed in polynomial order time by
  employing a dynamic programming algorithm (See, e.g., Durbin {\it et al.} \cite{durbin1998biological}.
  Like a BPPM for an RNA sequence, an APPM can be easily visualized.
  %
  \begin{example}[Leaf splitting probabilities]\label{eg:sp}
    A leaf splitting probability, $p_{X,Y}$, for a leaf set $S$ 
    gives the marginal probability that there exists a partition $(X,Y)$ of $S$ 
    (i.e., $X\cup Y=S$ and $X \cap Y = \emptyset$) formed by cutting an internal edge 
    in a phylogenetic tree:
    \begin{align}
      p_{X,Y}=\sum_{\theta\in \mathcal{C}(X,Y)} p(\theta| S)
    \end{align}
    where $C(X,Y)$ is the set of 
    phylogenetic tree topologies that can be split into $X$ and $Y$ by cutting one edge in $T$, and
    $p(\theta|S)$ is a probability distribution of phylogenetic tree topologies, such as the one given in \cite{pmid11524383}.
  \end{example}

  Unlike base-paring probabilities or aligned-pairing probabilities, 
  it is difficult to visualize the complete set of splitting probabilities 
  because it cannot be represented as a ``matrix''. 
  The following is a non-trivial visualization of phylogenetic trees, with leaf splitting
  probabilities.
  %
  \begin{example}[Centroid Wheel Tree (CWT) \cite{pmid20817714}]\label{eg:cwt}
    For phylogenetic tree estimations,
    a Centroid Wheel Tree (CWT) \cite{pmid20817714} provides 
    a novel visualization of a phylogenetic tree topology
    with marginal probabilities (leaf splitting probabilities; Example~\ref{eg:sp}).
    See Figure~\ref{fig:eg_cwt} for an example.
    See \cite{pmid20817714} for the detailed definition and algorithms for computing a CWT.
    A CWT can be computed through a Web site (http://cwt.cb.k.u-tokyo.ac.jp/).
  \end{example}
  %
  \begin{figure}[t]
    \centerline{
      \includegraphics[width=0.5\textwidth]{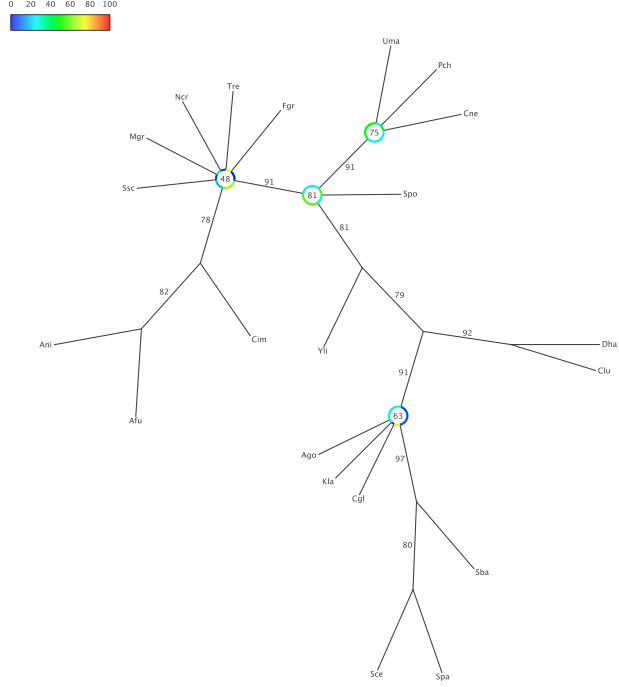}
    }\caption{\label{fig:eg_cwt}\small
      An example of a Centroid Wheel Tree (CWT) computed using the Web site: 
      http://cwt.cb.k.u-tokyo.ac.jp/.
      In this representation, leaf splitting probabilities are shown next to each edge.
    }
  \end{figure}

  \section{What should we do when {\em one} prediction is required?}\label{sec:one_solution}
  %
  Although the approaches described in the previous section are good ways to
  handle { uncertainty}, it is often necessary to predict only 
  {\em one} solution.

  \subsection{Maximum likelihood estimator and its drawback}

  A straight-forward approach to make a prediction for a given probability distribution, 
  $p(y|D)$, on the predictive space $Y$ is to find the solution $\hat y$ 
  with the maximum probability:
  \begin{align}
    \hat y = \argmax_{y\in Y} p(y|D).\label{eq:ML}
  \end{align}
  This approach is known as maximum likelihood (ML) estimation.
  Many existing tools or algorithms employ ML estimators; 
  however, recent studies have indicated that ML estimators 
  are not always superior estimators e.g. \cite{pmid18305160,pmid21365017}.
  The following is a typical example in which
  the ML estimator is inappropriate.
  %
  \begin{example}[Carvalho and Lawrence \cite{pmid18305160}]\label{eg:CL}
    The predictive space is the $n$-dimensional binary space $Y=\{0,1\}^n$, in which
    similar vectors in $Y$ represent similar predictions.
    We introduce a probability distribution on $Y=\{0,1\}^n$ as follows:
    \begin{align*}
      p(\theta|D)=\left\{
      \begin{array}{ll}
        p_1:=\frac{1}{n+3} & \mbox{if } \theta\in 
        \mathcal{S}:=\left\{(x_1,\ldots,x_n)| x_k \in \{0,1\}, \sum_{k=1}^n x_k \le 1 \right\}\\
        p_2:=\frac{2}{n+3} & \mbox{if } \theta=\theta^1:=(1,1,\ldots,1)\\
        0 & \mbox{otherwise}
      \end{array} 
      \right.
    \end{align*}
    (Clearly, $\sum_{\theta\in Y} p(\theta|D)=1$.) Then, the maximum likelihood estimator gives 
    the solution $\theta^1=(1,1,\ldots,1)$. 
    However, it can be seen that
    \begin{align*}
      \sum_{\theta: H(\theta^0, \theta)\le 1} p(\theta|D) = \frac{n+1}{n+3}
    \end{align*}
    where $\theta^0$ is the zero vector in $\{0,1\}^n$ (i.e., all elements in the vector are equal to 0) and $H(\cdot,\cdot)$
    is the Hamming distance.
    This means that, when $n=997$, the sum of the probabilities solutions around (similar to) $\theta^0$,
    for which the Hamming distance from $\theta^0$ is less than or equal to 1, is 0.998, 
    while the probability of
    the ML estimation (i.e., $\theta^1$) is only 0.002 (Figure~\ref{fig:CL}), 
    indicating that $\theta^0$ is a much better estimation than the ML estimation $\theta^1$.
    Note that the ML estimation $\theta^{1}$ is a long way from the solution $\theta^0$ 
    because $H(\theta^0,\theta^{1})=n$.
  \end{example}
  %
  \begin{figure}[t]
    \centerline{
      \includegraphics[width=0.75\textwidth]{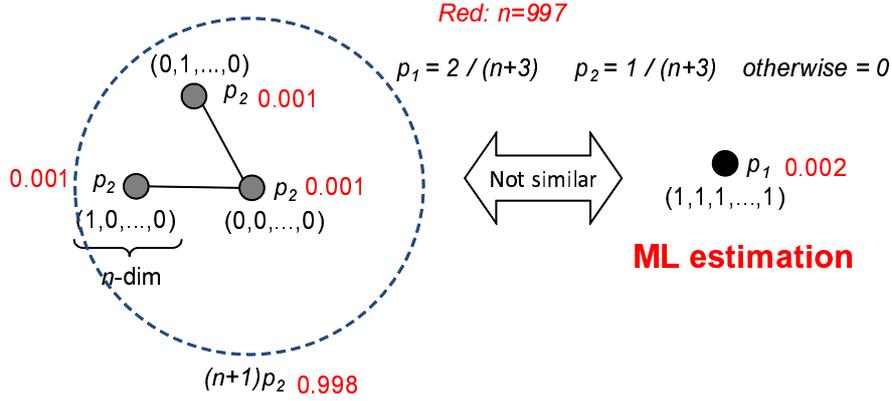}
    }\caption{\label{fig:CL}\small
      An example of Carvalho and Lawrence \cite{pmid18305160},
      indicating that the ML estimation may not be very good. 
      In this example, the predictive space is the $n$-dimensional binary space, $Y=\{0,1\}^n$, and we assume that similar binary vectors give similar predictions.
      Probabilities of each solution are shown in the figure (see the main text for the details) and edges are shown between solutions whose Hamming distance is 1
      The ML estimation is $\theta^1$ but the sum of probabilities around $\theta^0$ is much
      larger than the probability of $\theta^1$. See Example~\ref{eg:CL} in the main text for the details.
    }
  \end{figure}

  It is worth noting that several bioinformatics problems 
  (including 
  RNA secondary structure prediction,
  RNA-RNA interaction prediction,
  sequence alignment, 
  and
  phylogenetic tree estimation) can be formulated as a point estimation
  problem in (a subset of) binary spaces $Y=\{0,1\}^n$ with large $n$ \cite{pmid21365017}.
  Hence, a similar situation to the one in the above (artificial) example 
  could occur in {\em real} problems in bioinformatics.
  {%
    (For example, in Figure~\ref{fig:MDS_sfold}, the clover-leaf structure of tRNA 
    does not lie in the 1st rank cluster.)
  }

  \subsection{Maximum expected gain/accuracy (MEG/MEA) estimators}\label{sec:meg}
  %
  Because ML estimators are not always good estimators, as shown in Example~\ref{eg:CL}, 
  we will introduce alternatives to ML estimators in this section, 
  on the basis of given accuracy/evaluation measures for the prediction 
  problem.
  %
  \subsubsection{Definition}
  %
  Given a probability distribution $p(\theta|D)$ on a predictive space $Y$,
  the {\em maximum expected gain (MEG) estimator} is defined 
  by
  \begin{align}
    \hat y = \mathop{\mathrm{arg\ max}}_{y \in Y} \sum_{\theta\in Y} G(\theta,y) p(\theta|D)\label{eq:MEG}
  \end{align}
  where $G(\theta,y)$ is called the gain function, which returns the similarity between two solutions in $Y$.
  When the gain function is designed according to an accuracy or evaluation measure
  for the target problem, in which $y$ and $\theta$ are considered as a prediction and reference, respectively, 
  the estimator is often called a {\em maximum expected accuracy (MEA) estimator} 
  \cite{pmid16873527,pmid19095700,pmid22313125}%
  \footnote{%
    When this estimator is called a ``maximum expected accuracy'' (MEA) estimator, 
    $G(\theta, y)$ in Eq.~(\ref{eq:MEG}) is equal to an accuracy measure 
    (or is designed according to an accuracy measure) 
    for a reference $\theta$ and a prediction $y$. 
    This also implies that $p(\theta|D)$ 
    is considered to be a probability distribution of references, 
    which is misleading because $p(\theta|D)$ does not usually represent such a distribution. 
    In RNA secondary structure prediction, for example, 
    the McCaskill model provides not a probability distribution of reference secondary structures 
    but rather a full ensemble of possible secondary structures \cite{pmid1695107}.
  }.
  MEA estimators predict the solution by maximizing the expected accuracy when
  the solutions are distributed according to $p(\theta|D)$. These estimates are, therefore, appropriate for
  the given accuracy measure\footnote{%
    MEA estimators are closely related to posterior decoding algorithms (cf. Section~\ref{sec:post_dec}).
  }.
  %
  \subsubsection{ML estimators from the viewpoint of MEA}
  %
  Let us consider the ML estimator from the viewpoint of MEA.
  The ML estimator of Eq.~(\ref{eq:ML}) can be represented as
  \begin{align*}
    \hat y  &= \mathop{\mathrm{arg\ max}}_{y \in Y} p(y|D)\\
    &= \mathop{\mathrm{arg\ max}}_{y \in Y} \sum_{\theta\in Y} \delta(\theta,y) p(\theta|D)
  \end{align*}
  where 
  $\delta(\theta,y)$ is the Kronecker delta function 
  that returns 1 only if $\theta=y$.
  This means the ML estimator is a kind of MEG/MEA estimator 
  in which the gain (accuracy) function is the delta function.
  However, the delta function is rarely employed as an accuracy measure for most problems.
  In RNA secondary structure predictions, for instance, using the delta function as an
  accuracy measure would mean that a predicted secondary structure must be
  exactly same as the reference structure; this seems too strict.
  Instead, the evaluation is usually 
  {based on a comparison of base-pairs in predicted and reference secondary structures
    (Supplementary Section~\ref{sec:acc_meas_rna})}.

  Note that the delta function $\delta(\theta,y)$ is much stricter 
  than the accuracy measures in { Supplementary Section~\ref{sec:acc_meas_rna}}. 
  The MFE structure is equivalent to the ML estimator with the probability distribution of Eq.~(\ref{eq:BD})
  and therefore the MFE structure is not optimal to 
  the accuracy measures above.
  How do we design an MEA estimator for which it is appropriate to use those accuracy measures?

  \subsubsection{Example: RNA secondary structure prediction (I)}\label{sec:g-centroid}
  %
  In RNA secondary structure predictions, 
  more true predictions (TP/TN) and 
  fewer false predictions (FP/FN) of base-pairs should occur in a 
  predicted secondary structure (cf. {Section~\ref{sec:acc_meas_rna}}).
  Based on the principle of the MEA estimator, 
  the following MEG estimators (called generalized centroid estimators) have been introduced \cite{pmid19095700}:
  \begin{align*}
    \hat y = \mathop{\mathrm{arg\ max}}_{y \in \mathcal{S}(x)} \sum_{\theta\in \mathcal{S}(x)} G(\theta,y) p(\theta|x)
  \end{align*}
  where
  \begin{align*}
    {G}(\theta, y) = \alpha_1 \cdot TP + \alpha_2 \cdot TN - \alpha_3 \cdot FP - \alpha_4 \cdot FN
  \end{align*}
  with the $\alpha_k$ representing arbitrary positive constants, and
  $p(\theta|x)$ is a probability distribution on a set $\mathcal{S}(x)$ 
  of possible secondary structures of RNA sequence $x$; 
  {TP, TN, FP and FN are defined in Supplementary Section~\ref{sec:acc_meas_rna}.}
  Hamada {\it et al.}~\cite{pmid19095700} showed 
  that the number of parameters can be reduced to only one,
  because the MEA estimator with this gain function is equivalent to the MEG estimator with the gain function
  \begin{align}
    G^{(c)}_\gamma(\theta,y) &= \gamma \cdot TP + TN \label{eq:gain_centroid},
  \end{align}
  where $\gamma > 0$. 
  When $\gamma=1$, the estimator is equivalent to the centroid estimator \cite{pmid18305160}, so
  the estimator is called the generalized centroid estimator or $\gamma$-centroid estimator.
  By using the parameter $\gamma$, users can adjust the balance between 
  sensitivity and PPV (cf. Eqs.~(\ref{eq:SEN}) and (\ref{eq:PPV})).
  
  CONTRAfold \cite{pmid16873527} employs a different MEA estimator, but 
  Hamada {\it et al.}~\cite{pmid19095700} theoretically showed that the gain function used 
  in CONTRAfold includes a term that is biased with respect to the accuracy measure of RNA secondary structure prediction.
  Computational experiments reported in \cite{pmid19095700} 
  support this theoretical result because
  they obtained consistent results for 
  three different probabilistic models 
  for RNA secondary structures.
  This example indicates the importance of 
  designing the gain function appropriately.

  \subsubsection{Example: RNA secondary structure prediction (II)}\label{sec:rna_sec_2}
  %
  In the previous section, a linear combination of TP, TN, FP and FN is utilized as
  a gain function.
  Here, we try to optimize the accuracy given by the measure MCC (Eq.~(\ref{eq:MCC}) {in Supplementary Section~\ref{sec:acc_meas_rna}}) directly 
  in the MEA estimators. Thus, the MEA estimator
  \begin{align}
    \hat y =\mathop{\mathrm{arg\ max}}_{y\in \mathcal{S}(x)} \sum_{\theta\in\mathcal{S}(x)}\mbox{MCC}(\theta,y)p(\theta|x).\label{eq:mea_mcc}
  \end{align}
  is introduced.
  Unfortunately, it is difficult to compute this estimator efficiently, because
  neither the expectation of MCC nor the
  arg max operation in Eq.~(\ref{eq:mea_mcc}) can be computed efficiently.
  In \cite{pmid21118522} the authors addressed this problem by proposing 
  approximate estimators 
  (called maximum {\em pseudo}-expected accuracy estimators) as follows:
  \begin{align}
    \hat y = \mathop{\mathrm{argmax}}_{y\in\mathcal{S}(x)} \widehat{\mbox{MCC}}^0(y),\label{eq:approx_mcc}
  \end{align}
  where
  \begin{align}
    \widehat{\mbox{MCC}}^0(y)=
    \frac{\widehat{\mbox{TP}}\cdot 
      \widehat{\mbox{TN}}-\widehat{\mbox{FP}}\cdot 
      \widehat{\mbox{FN}}}{\sqrt{(\widehat{\mbox{TP}}+
        \widehat{\mbox{FP}})
        (\widehat{\mbox{TP}}+\widehat{\mbox{FN}})
        (\widehat{\mbox{TN}}+\widehat{\mbox{FP}})
        (\widehat{\mbox{TN}}+\widehat{\mbox{FN}})}}.\label{eq:pmcc}
  \end{align}
  and $\widehat{\mbox{TP}}$, $\widehat{\mbox{TN}}$, $\widehat{\mbox{FP}}$ and $\widehat{\mbox{FN}}$ 
  are, respectively, the expected values of TP, TN, FP and FN (under a given probability distribution $p(\theta|x)$).
  In contrast to the expected value of MCC, $\widehat{\mbox{MCC}}^0(y)$ can be computed efficiently
  from the base-pairing probability matrix.
  Stochastic sampling is employed to approximate the ``argmax'' operation in Eq.~(\ref{eq:approx_mcc}). 

  {
    It should be emphasized that the above estimators can be adapted to utilize more general accuracy measures.
    See Supplementary Section~\ref{sec:max_pseudo_acc} for details.
  }
  %
  \subsubsection{Example: Sequence alignments}
  %
  The MEA estimators introduced in the previous sections (for RNA secondary structure predictions)
  can be directly applied to pairwise sequence  alignments,
  where TP, TN, FP and FN are computed with respect to 
  bases (nucleotides or amino acids) that are aligned (and un-aligned) between
  two sequences, and therefore the $\gamma$-centroid estimator is expected to
  produce more correctly aligned-bases in the predicted alignment.
  Actually, computational experiments in Frith {\it et al.}~\cite{pmid20144198}
  indicated that, compared to the conventional alignment method 
  (i.e., maximizing an alignment score that corresponds to the ML estimation of 
  pairwise alignment), the $\gamma$-centroid alignments substantially
  reduce the number falsely aligned bases (i.e., FP) in return for the sacrifice 
  of a slight reduction in the number of correctly
  aligned bases (i.e., TP). 
  %
  \subsubsection{Example: Gene predictions}
  %
  In gene prediction \cite{pmid12209144,pmid22510764}, 
  the accurate prediction of a {\em boundary} (e.g., an exon-intron boundary) in the predicted gene 
  is important.
  Sensitivity and specificity have been used for evaluation at the gene and exon levels \cite{pmid18096039,pmid22510764}.
  In order to design a predictor that is suited to these accuracy measures,
  the number of correctly predicted boundaries and un-boundaries can be utilized
  in the gain function, and the corresponding MEA estimator is called a maximum expected boundary accuracy
  estimator.
  {%
    Like MEA estimators in RNA secondary structure prediction, the prediction can be efficiently conducted 
    by utilizing a dynamic programming approach.
  }
  See Gross {\it et al.}~\cite{pmid18096039} for the details.
  In their computational experiments, the MEA estimator based on boundary accuracy measures 
  was superior to existing gene prediction methods.

  \subsubsection{Other examples}
  %
  In addition to the above examples,
  many algorithms in bioinformatics can be classified, from the viewpoint of MEA/MEG,
  with respect to gain function and predictive space. 
  See \cite{pmid22313125} for a review of MEA estimators.

  \subsection{Point estimation with additional information}\label{sec:pe_plus}
  %
  Although well-designed MEG/MEA estimators would 
  provide a better single solution 
  that is more appropriate to a given evaluation measure 
  than ML estimators, 
  the probability of the prediction is still low. In fact, it is lower than the probability of the ML estimation.
  In real applications, it can be useful to provide a single solution along with
  additional information as we now show\footnote{%
    Multiple pieces of information can be simultaneously utilized as additional information for a point estimation. 
  }.

  \subsubsection{Point estimation with marginal probabilities}\label{sec:PE_marg}
  %
  Marginal probabilities (cf. Section~\ref{sec:marg}) can be often used as 
  additional information for a point estimation as demonstrated by the following examples.
  %
  \begin{example}[RNA secondary structure with base-pairing probabilities (BPPs)]\label{eg:sec_w_prob}
    Base-pairing probabilities (Example~\ref{eg:bpp}) can be used as additional information 
    for a point estimation of RNA secondary structure
    as shown in Figure~\ref{fig:tRNA-cf}.
    This kind of prediction is available by using
    the CentroidFold Web server \cite{pmid19435882} and the RNAfold Web server~\cite{pmid22115189}.
  \end{example}
  %
  \begin{example}[Pairwise alignment with aligned-pairing probabilities (APPs)]\label{eg:alingment_w_prob}
    Aligned-pairing probabilities (cf. Example~\ref{eg:app}) 
    can be used as additional information for the point estimation of pairwise 
    alignments
    (Figure~\ref{fig:alignment-last}).
    The LAST web server (http://last.cbrc.jp/) returns a pairwise alignment with
    aligned probabilities.
    Also, FSA (Fast Statistical Alignment) \cite{pmid19478997} computes 
    marginal probabilities including {gap} probabilities\footnote{%
      Gap probabilities are the marginal probabilities that a specific position corresponds 
      to a gap, which can be computed by a dynamic programming algorithm.
    }.
  \end{example}

  Kim {\it et al.}~\cite{pmid21576232} have extended aligned-pairing probabilities 
  to the case of multiple alignments:
  probabilistic sampling-based alignment reliability (PSAR) calculates 
  the reliability of each column in a given multiple alignment.

  %
  \begin{figure}[t]
    \centerline{
      \includegraphics[width=0.3\textwidth]{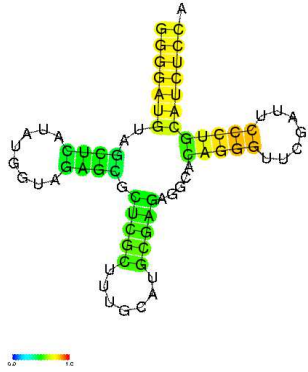}
    }
    \caption{\label{fig:tRNA-cf}\small
      Secondary structure of a tRNA sequence with base-pairing probabilities 
      produced by the CentroidFold Web Server (http://www.ncrna.org/centroidfold) \cite{pmid19435882}.
      Base-pairs with warmer colors have higher base-pairing probabilities.
    }
  \end{figure}
  %
  \begin{figure}[t]
    \centerline{
      \includegraphics[width=0.95\textwidth]{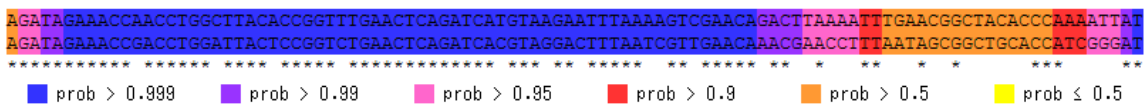}
    }
    \caption{\label{fig:alignment-last}\small
      A pairwise alignment with aligned-pairing probabilities 
      produced by the LAST Web Server (http://last.cbrc.jp/) \cite{pmid20144198}.
      Aligned-pairs with colder colors have higher aligned pairing probabilities.
    }
  \end{figure}

  \subsubsection{Point estimation with credibility limits}\label{sec:PE_CL}
  %
  The {\em credibility limit} \cite{pmid18464927} of $\alpha$ ($0\le \alpha \le 1$) 
  provides a {\em global} measure of the reliability of a point estimation, 
  which is computed in the following way.
  \begin{enumerate}
  \item A probability distribution $p(\theta|D)$ for a predictive space $Y$ is given
    (e.g., a probability distribution $p(\theta|x)$ for secondary structures of RNA sequence $x$ is given).
  \item A distance between two solutions is defined. 
    (For example, the base-pair distance 
    can be used for RNA secondary structure predictions.)
  \item Fix an arbitrary solution $y\in Y$ that you are interested in.
    (For example, fix the MFE structure of the RNA sequence $x$.)
  \item Compute the minimum distance $d$ for which $100\times \alpha$ \% {of the accumulated probabilities} of all solutions are included
    in the set of solutions whose distance from $y$ is less than $d$. This distance is called the credibility limit of
    $\alpha$.
  \end{enumerate}
  The authors \cite{pmid18464927} have constructed an algorithm to compute credibility limits,
  based on stochastic sampling of RNA secondary structures or alignments (Section~\ref{sec:suboptimal}). 
  Note, however, that credibility limits for RNA secondary structures and alignments 
  can be computed by using dynamic programming
  without using sampling techniques \cite{pmid19119992}, because
  the complete distribution from a fixed solution can be computed by using dynamic 
  programming (see Section~\ref{sec:dist_sol}).
  %
  \section{Handling uncertainty to develop complex pipelines}\label{sec:pipeline} 
  %
  \subsection{General strategies}\label{sec:general_strategy}
  %
  When developing complex algorithms or pipelines 
  in which 
  two or more prediction methods are involved, 
  the uncertainty described in this review should be handled carefully, 
  especially for the {\em intermediate} prediction method(s) in the pipeline.
  For example,
  let us consider the following pipeline, which is based on two prediction methods
  (with the example of phylogenetic tree estimation from unaligned sequences in parentheses; 
  see Figure~\ref{fig:prob_tree}):
  \begin{enumerate}
  \item Obtain data $D$ (e.g., $D$ is a set of {\em unaligned} sequences $S$).
  \item Predict an intermediate solution $\theta' \in Y'$ from $D$   
    (e.g., predict a multiple sequence alignment (MSA) $A$ of $S$).
  \item Predict the final solution $\theta \in Y$, based on $\theta'$ 
    (e.g., predict a phylogenetic tree $T$ based on the multiple alignment $A$)
  \end{enumerate}
  %
  Given the uncertainty of the intermediate estimation problem 
  (in step 2 above), 
  predicting a single solution $\theta'$ should be avoided
  because point estimation is unreliable.
  Actually, as discussed in Section~\ref{sec:intro}, the prediction of phylogenetic trees
  is greatly affected by the uncertainty of multiple alignments \cite{pmid18218900}.
  To handle the uncertainty, it is ideal to consider a joint distribution $p(\theta, \theta'|D)$ on 
  $Y\times Y'$. For instance, $p(\theta,\theta'|D)$ could be a joint distribution for a multiple sequence alignment (MSA) and a phylogenetic tree.
  Then the distribution is marginalized onto $Y$:
  \begin{align}
    p(\theta|D) = \sum_{\theta'\in Y'} p(\theta, \theta'|D)\;.\label{eq:joint}
  \end{align}
  This marginal distribution includes the uncertainty of the intermediate solution $y'$.
  From this marginal distribution, we can predict either one final solution, 
  by using, for example, ML/MEA/MEG estimators (Section~\ref{sec:one_solution}),
  or several solutions (see Section~\ref{sec:giveup}).
  However, considering joint distributions (such as Eq.~(\ref{eq:joint})) often
  leads to huge computational cost, and
  an approximation could be useful to obtain efficient estimators as shown 
  in the following examples.
  %
  \begin{figure}[t]
    \centerline{
      \includegraphics[width=0.4\textwidth]{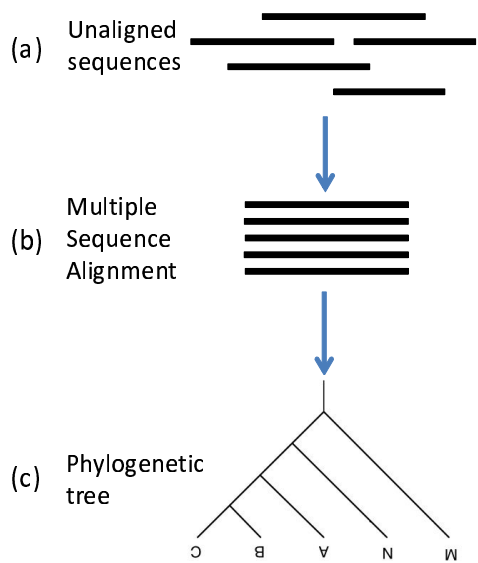}
    }\caption{\label{fig:prob_tree}\small
      A standard method to estimate a phylogenetic tree (topology) 
      from a set of {\em unaligned} sequences, 
      in which (b) a multiple sequence alignment (MSA) of sequences is predicted by using 
      a multiple-aligner such as ClustalW or MAFFT \cite{pmid23329690}, 
      and then (c) a phylogenetic tree is predicted from the MSA.
      {\em Point} estimation of an MSA (i.e., step (b)) influences uncertainty and should be avoided 
      if possible.
    }
  \end{figure}

  \subsection{Example: Prediction of RNA secondary structure by 
    employing homologous sequence information}\label{sec:prob_hom}
  %
  In this section, we consider the problem of predicting an RNA secondary structure
  by combining homologous sequence information,
  where
  the input is the target RNA sequence $t$ 
  and a set of its homologous sequences $H$ (see Figure~\ref{fig:prob_hom}).
  In this problem, we assume that 
  the target sequence $t$ and each sequence in $H$ share a consensus structure, so the information
  from the homologous sequences would improve the accuracy of the RNA secondary structure
  prediction of the target sequence.

  {One approach} to this problem is the following (Figure~\ref{fig:prob_hom})\footnote{%
    {Covariance models \cite{pmid8029015} can also be used to solve the problem.}
  }.
  \begin{enumerate}
  \item A multiple sequence alignment for input RNA sequences 
    (both target and homologous sequences) 
    is computed by utilizing a multiple aligner for RNA sequences 
    (such as CentroidAlign \cite{pmid19808876}, Locarna \cite{pmid17432929}  and MAFFT \cite{pmid23329690}).
  \item A common/consensus secondary structure for the multiple alignment 
    is computed by using, e.g., CentroidAlifold \cite{pmid20843778} and RNAalifold \cite{pmid19014431}.
  \item The secondary structure of the target RNA sequence is computed from 
    the predicted common secondary structure by mapping the common structure to the target RNA sequence.
  \end{enumerate}
  %
  Note that intermediate point estimations are employed twice (in steps 1 and 2) in 
  the above procedure. The results will be affected by the uncertainty of 
  these point estimations, which should therefore be avoided (Section~\ref{sec:general_strategy}).

  There is a probabilistic model for {\em structural} alignments 
  of RNA sequences (known as the Sankoff model \cite{sankoff1985}), 
  in which both consensus secondary structure and conventional alignments 
  are considered simultaneously\footnote{%
    Strictly speaking, Sankoff \cite{sankoff1985} 
    did not use a probabilistic model for structural alignments.
    However, one can be easily introduced by using a similar approach 
    to that of McCaskill \cite{pmid1695107}.
  }.
  As an alternative to point estimation,
  it is better to obtain a distribution of 
  RNA secondary structures of the target sequence $t$ by 
  marginalizing (and averaging)
  the distribution of structural alignments, and then predict a secondary structure 
  for the target 
  by using MEG/MEA estimators or the ML estimator with respect to the marginal probability distribution.
  However, this approach entails huge computational cost for prediction 
  ($O(L^6)$ where $L$ is the length of RNA sequences).
  Hamada {\it et al.}~\cite{pmid19478007} approximate 
  this approach by factorizing the distribution of structural alignments into 
  a distribution of secondary structures and a distribution 
  of usual alignments\footnote{This algorithm is implemented in 
    the CentroidHomfold software \cite{pmid19478007}.}.
  Furthermore, they showed that the approach in Figure~\ref{fig:prob_hom} 
  (in which the { uncertainty} of secondary structures and alignments is not considered) 
  was consistently
  worse than their method in which the {uncertainty} of alignments and secondary structures
  is considered. 
  See \cite{pmid19478007} for the details. 

  A similar type of technique was utilized for sequence alignments of 
  RNA sequences \cite{pmid19808876}.
  Moreover, a more general form of estimator was introduced in \cite{pmid21365017}.
  %
  \begin{figure}[t]
    \centerline{
      \includegraphics[width=0.5\textwidth]{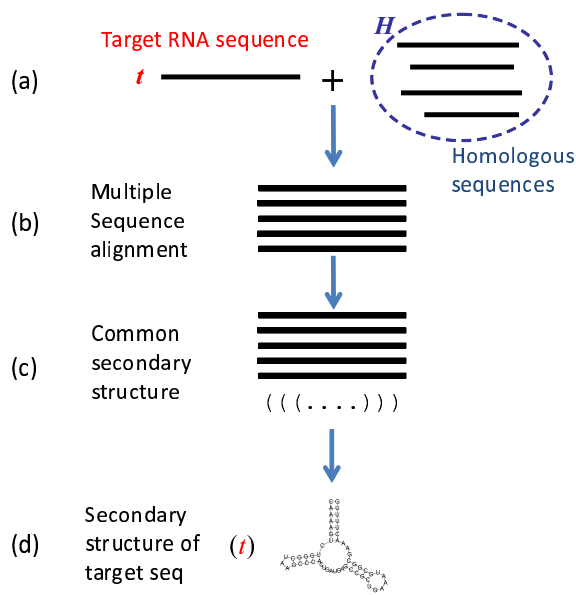}
    }\caption{\label{fig:prob_hom}\small
      RNA secondary structure prediction using homologous sequence information.
      A typical method used to solve this problem is shown, in which two (intermediate) point estimations are
      made.
      (a) The input is a target RNA sequence $t$ (whose secondary structure is to be predicted) and
      a set of homologous sequences $H$. The output is an RNA secondary structure for the 
      target sequence.
      In the problem, we assume that
      the target sequence $t$ and each sequence in $H$ share a consensus structure.
      In the conventional method, (b) a multiple sequence alignment of the target and homologous
      sequences is predicted; then (c) a common (consensus) secondary structure 
      of the MSA is predicted;
      finally, (d) the secondary structure of the target sequence is predicted by using
      the common secondary structure.
      A more sophisticated approach avoiding point estimations is described in the
      main text (Section~\ref{sec:prob_hom}).
    }
  \end{figure}

  {%
    \subsection{Example: Variant detection with NGS technologies}
    %
    In the following pipeline, I try to detect variants of genomes, 
    using next generation sequencing (NGS) technologies.
    \begin{enumerate}
    \item 
      Determine bases (by ``base-calling'' \cite{pmid21245079}) from intensity data produced by sequencers 
      (e.g. Illumina, 454, SOLiD and PacBio RS), 
      and obtain a set of reads (i.e., fragmented short sequences). 
    \item
      Map (align) the reads to reference genomes by utilizing, 
      e.g., BWA \cite{pmid19451168}, bowtie \cite{pmid22388286} or LAST \cite{pmid23413433}.
    \item 
      Predict single nucleotide polymorphisms (SNPs) and insertions and deletions (INDELs) 
      from the mapped reads obtained in Step 2, 
      using tools such as 
      SAMtools \cite{pmid19505943} or VarScan~\cite{pmid22300766}.
    \end{enumerate}

    In this pipeline, several point estimations are employed:
    Step 1 includes the uncertainty of called bases.
    By retaining the information about the uncertainty of base-calling, most current sequencers
    produce reads  with quality scores in the FASTQ  format \cite{pmid20015970}.
    Step 2 includes uncertainty with respect to the locations of mapped reads 
    and the detailed alignment between reads and reference genomes.
    To the best of my knowledge, there is no method that handles these uncertainties completely,
    although the approach of {\em probabilistic} alignments with quality scores \cite{pmid21976422} 
    partially handles both uncertainties\footnote{%
      The uncertainty arising from locations of mapped reads cannot be resolved completely 
      even if probabilistic alignments with quality scores are utilized 
      in the pipeline.
    }.
  }

  \section{Discussion}\label{sec:discussion}

  \subsection{Usefulness of marginal probabilities}\label{sec:useful_marg}
  %
  \subsubsection{Analyses using marginal probabilities}\label{sec:useful_marg_anal}

  Due to the uncertainty of single solutions that has been described in this article,
  considering an ensemble of solutions is useful in the analysis of biological data in bioinformatics.
  In particular, marginal probabilities (cf. Section~\ref{sec:marg}) 
  are often employed in the analyses, because they take into account information 
  about the entire ensemble of solutions.

  \begin{itemize}
  \item 
    In Adachi {\it et al.}~\cite{pmid21524680}, 
    the analysis of the base-paring probability 
    matrix of an RNA aptamer (bound to a cytokine) 
    suggested that a stem involved in the aptamer was unstable and 
    there was a possibility that it could form several different structures.
    This was confirmed by biochemical experiments (see Figure~\ref{fig:apt_bppm}). 
    It should be emphasized that we could not have obtained this conclusion 
    if only a single predicted structure (e.g., MFE structure) had been considered.
    \begin{figure}[t]
      \centerline{
        \includegraphics[width=0.6\textwidth]{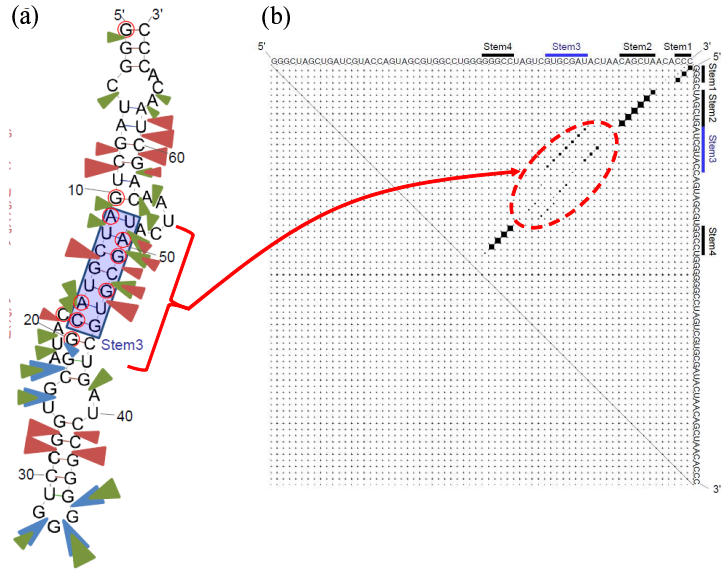}
      }\caption{\label{fig:apt_bppm}\small
        RNase probing analysis of AptAF42dope1. 
        (a) Fluorescent images of the cleavage products of 50-FAM labeled aptamer 
        with RNase T1, S1 nuclease and RNase V1. 
        (b) Secondary structure of an RNA aptamer predicted by CentroidFold \cite{pmid19435882}, 
        and a base-pairing probability matrix of the aptamer \cite{pmid21524680}. 
        The base-pairing probabilities in the red circle correspond to an unstable stem.
        This figure is taken from Figure 4 and Figure 6
        in Adachi {\it et al.}~\cite{pmid21524680}.
      }
    \end{figure}
    %
  \item 
    In Halvorsen {\it et al.}~\cite{pmid20808897}, 
    the authors analyzed the conformation change 
    of RNA secondary structures, 
    cased by a single nucleotide polymorphism (SNP), 
    by utilizing the base-pairing probabilities. 
    They reported seven disease states and phenotypes in which two or more
    associated SNPs were found to alter the structural ensemble of
    the RNA (see Table~1 in \cite{pmid20808897} for the details).
  \item 
    Iwakiri {\it et al.} \cite{iwakiri-hamada} systematically analyzed 
    base-paring probabilities for paired and unpaired nucleotides of RNA secondary structures 
    involved in known protein-RNA complexes taken from the Protein Data Bank (PDB) \cite{pmid10592235}.
    Their analyses lead to novel findings that could not be found 
    if only the 
    snapshots of RNA secondary structures in the PDB are considered.
  \end{itemize}

  \subsubsection{Algorithms using marginal probabilities: posterior decoding}\label{sec:post_dec}

  It should be also remarked that marginal probabilities 
  are employed to construct algorithms, 
  and such an approach
  is called {\em posterior decoding}.
  Posterior decoding algorithms are widely utilized in bioinformatics:
  see \cite{pmid8771180,pmid9773345,pmid10383470,pmid21976422,pmid18073381,pmid16351738,pmid17068081,pmid20576627,pmid18420654,pmid15804354}, 
  for example~\footnote
  { From a historical viewpoint, a posterior decoding was originally proposed in 
    Miyazawa \cite{pmid8771180} for alignments, later
    adopted for HMMs by Holmes and Durbin \cite{pmid9773345},
    and first introduced into RNA secondary structure predictions 
    by Knudsen and Hein \cite{pmid10383470}.
  }.
  Actually, most MEG/MEA estimators (Section~\ref{sec:meg}) 
  lead to posterior decoding algorithms, because the final algorithm 
  is based on only marginal probabilities. For example, the final algorithm in
  the $\gamma$-centroid estimators for RNA secondary structure predictions 
  (cf. Section~\ref{sec:g-centroid})
  is based on the base-pairing probability matrix (BPPM) of the RNA sequence 
  (see \cite{pmid19095700} for the detailed algorithm).

  \subsection{Uncertainty in hypothesis testing}
  
  In several problems in bioinformatics such as homology search \cite{pmid20180275},
  a hypothesis testing approach is frequently employed, 
  in which the log odds score of two probabilities relative to 
  a null model is computed instead of the maximum score of the solution. 
  For instance,
  in homology search, the following (Viterbi) score is utilized:
  \begin{align}
    V = \log \frac{\max_z p(x,z|H)}{p(x|R)}
  \end{align}
  where $x$ is a target sequence (we would like to judge whether this sequence 
  is homologous to a query sequence or not), 
  $R$ is a random model 
  (e.g., a one-sate HMM), and $z$ is an alignment 
  between the target and query sequences.
  The score might be affected by uncertainty because it uses only the maximum score of the
  optimal alignment.
  To address this issue, Eddy \cite{pmid18516236} proposed the use of forward scores, 
  \begin{align}
    F = \log \frac{\sum_z p(x,z|H)}{p(x|R)},
  \end{align}
  in homology searches of biological sequences.
  In the forward score,
  the {\em sum} of the probabilities of all solutions (alignments) are utilized, which
  mitigates the influence of the uncertainty \cite{pmid18516236}.

  \subsection{Uncertainty introduced by internal parameters in models}

  In the above, I have focused on the uncertainty of solutions 
  where the probabilistic model 
  (i.e., the probability distribution on the predictive (solution) space)
  is fixed.
  On the other hand, 
  a kind of uncertainty also appears 
  with respect to changes of internal parameters 
  in the probabilistic model 
  (e.g., a predicted sequence alignment might be substantially 
  changed by slight changes of the
  substitution matrix or gap costs in alignments). 
  Several interesting studies have been conducted to address this issue.

  In \cite{pmid21409513,pmid15534223,pmid16789815},
  parametric approaches are utilized to enumerate the optimal solutions
  for all possible changes of internal parameters,
  which
  enables the analysis of the parametric behavior of 
  {\it maximum  a posteriori} (MAP) inference computations.
  For example,
  a {\em parametric alignment}  procedure \cite{citeulike:4164936,pmid16789815} 
  can efficiently 
  find {\em all} the optimal alignments for all possible parameters 
  in a pair Hidden Markov Model (PHMM), 
  which might be one remedy for the uncertainty of solutions introduced 
  by parameter changes.
  It should be emphasized that the parametric methods can be explained 
  from a unified viewpoint by using a general mathematical theory called 
  {\em tropical geometry} \cite{pmid15534224}.

  \subsection{How to provide probability distributions for solutions}\label{sec:prob_model}
  %
  In this review, I have assumed that 
  there exists a probability distribution on a predictive space 
  in the estimation problem (Section~\ref{sec:purpose}).
  However, the existence of the probability distribution is not trivial 
  and research on the probability distribution (probabilistic model)
  for each problem 
  is also important, and  much research has been conducted in this area.
  For example,
  probabilistic models have been developed for
  RNA secondary structures 
  \cite{pmid1695107,pmid20940338,pmid16873527,DBLP:journals/jbcb/SatoHMAS10,pmid22194308},
  alignments \cite{pmid8771180,durbin1998biological,citeulike:1667683},
  gene prediction \cite{pmid18096039,pmid12538242},
  phylogenetic trees \cite{pmid11524383} and 
  RNA common secondary structures \cite{pmid19014431,pmid20843778}.

  \subsection{Future directions}
  %
  By using the approaches described in this review, 
  issues arising from the uncertainty of solutions can be addressed, 
  in part,
  in several areas of bioinformatics, including RNA secondary 
  structure predictions and sequence alignments. 
  However, it is obvious that uncertainty will raise further serious issues in other fields of
  bioinformatics.
  So, in the future, further general methods for fighting against uncertainty, 
  which are applicable to 
  various problems, should be developed: for example,  
  more efficient methods for visualizing or sampling solutions 
  in high-dimensional spaces,  
  methods that mitigate the uncertainly of point estimations, 
  and so forth. 
  
  \section{Conclusion}
  %
  In this review, I focused on the uncertainty of solutions 
  (i.e., the fact that the probability of any solution is quite low), 
  which often give rise to serious issues when developing algorithms 
  and analyzing biological data, and
  introduced several approaches to handle this problem appropriately.
  I have presented many actual examples in this review and,
  although most of the examples are related to 
  sequence analyses, such as RNA secondary structure predictions, 
  sequence alignments 
  and phylogenetic tree estimations, 
  I believe that the basic concepts and ideas described 
  in this review will be useful for other problems 
  in many areas of bioinformatics.

  \section*{Key Points}

  \begin{itemize}
  \item  Many bioinformatics problems, such as sequence alignment, gene prediction, 
    phylogenetic tree estimation and RNA secondary structure prediction, 
    are often affected by the ``uncertainty'' of a solution;
    that is, the probability of the solution is extremely small 
    { (e.g., less than $10^{-10}$)}. 
  \item In the analysis of biological data or the development of prediction algorithms, 
    this uncertainty should be handled carefully and appropriately. 
  \item This review provides several {approaches} to combat this uncertainty,  
    presenting examples in bioinformatics.
  \item {In particular, point estimations should be handled carefully when designing a pipeline.}
  \item The basic concepts and ideas described in this review will be useful in
    {various} areas of bioinformatics.
  \end{itemize}

  \section*{Biographical note}
  
  M.H. is an associate professor at
  Department of Computational Biology, the University of Tokyo.
  His background is pure mathematics and 
  his research interests include any mathematical aspects 
  of bioinformatics.

  \section*{Acknowledgement}
  %
  This work was supported in part by MEXT KAKENHI 
  (Grant-in-Aid for Young Scientists (A): 24680031).
  I am grateful to Dr.~Martin Frith, Dr.~Dave duVerle and Prof.~Kiyoshi Asai for valuable comments.
  Thanks are also due to three anonymous reviewers for useful suggestions, 
  which helped to improve the article.
  %
  \bibliographystyle{unsrt}
  \small \bibliography{main}
  
  \clearpage
  \appendix

  \setcounter{page}{1}

  \renewcommand{\thefigure}{S\arabic{figure}}
  \renewcommand{\thetable}{S\arabic{table}}
  \renewcommand{\theequation}{S\arabic{equation}}
  \setcounter{figure}{0}
  \setcounter{table}{0}
  \setcounter{equation}{0}
  \setcounter{footnote}{0}

  \section*{Supplementary Information}

  The IDs of references correspond to those of the main manuscript in the following.

  {
    \section{How to compute suboptimal solutions in Mfold}\label{sec:mfold}

    In this appendix, I briefly introduce an algorithm that predicts 
    suboptimal RNA secondary structures, which is described 
    in Zuker \cite{pmid2468181}\footnote{The algorithm is also described 
      on the Web: http://mfold.rna.albany.edu/doc/mfold-manual/node11.php}.

    Let $\Delta G$ be the minimum free energy (MFE) of secondary structures of 
    a given RNA sequence, 
    and let
    $\Delta G(i,j)$ be the minimum free energy of secondary structures 
    that contain the $(i,j)$ base-pair\footnote{%
      A base-pair formed from the $i$-th nucleotide and $j$-th nucleotide in the sequence.}.
    Then, the {\em energy dot plot} (Figure~\ref{fig:energy_dot_plot}) 
    is defined as a set of base-pairs as follows:
    \begin{align*}
      \mathcal{P}=\{(i,j)| \Delta G (i,j) \le \Delta G + \varepsilon\}
    \end{align*}
    where
    $\varepsilon$ is equal to  $\Delta G \times \frac{p}{100}$\footnote{The parameter $p$ is usually set to 5, i.e., a 
    5\% increase in the free energy is allowed.}.
    Then, Mfold computes a set of suboptimal secondary structures (denoted by $\mathcal{S}$) 
    as follows:
    \begin{enumerate}
    \item $\mathcal{S}$ is set to $\emptyset$.
    \item  $L$ is a list of base-pairs in $\mathcal{P}$, which is sorted in order of increasing $\Delta G(i,j)$.
    \item The base-pair $(i,j)$ at the top of list $L$ is selected.
    \item Predict the secondary structure $s$, with minimum free energy, 
      that contains the base-pair $(i,j)$.
    \item If $s$ contains at least $W$ base pairs that were not found in previous foldings (in $\mathcal{S}$)
      then make the replacement $\mathcal{S} \leftarrow \mathcal{S} \cup \{s\}$ and 
      remove the base-pairs of $s$ from the list $L$.
    \item Go to Step 2 and repeat the procedure until the maximum number 
      of suboptimal secondary structures (specified by a user) is reached or no base-pair is found in $L$.
    \end{enumerate}
  }

  \begin{figure}[ht]
    \centerline{
      \includegraphics[width=0.7\textwidth]{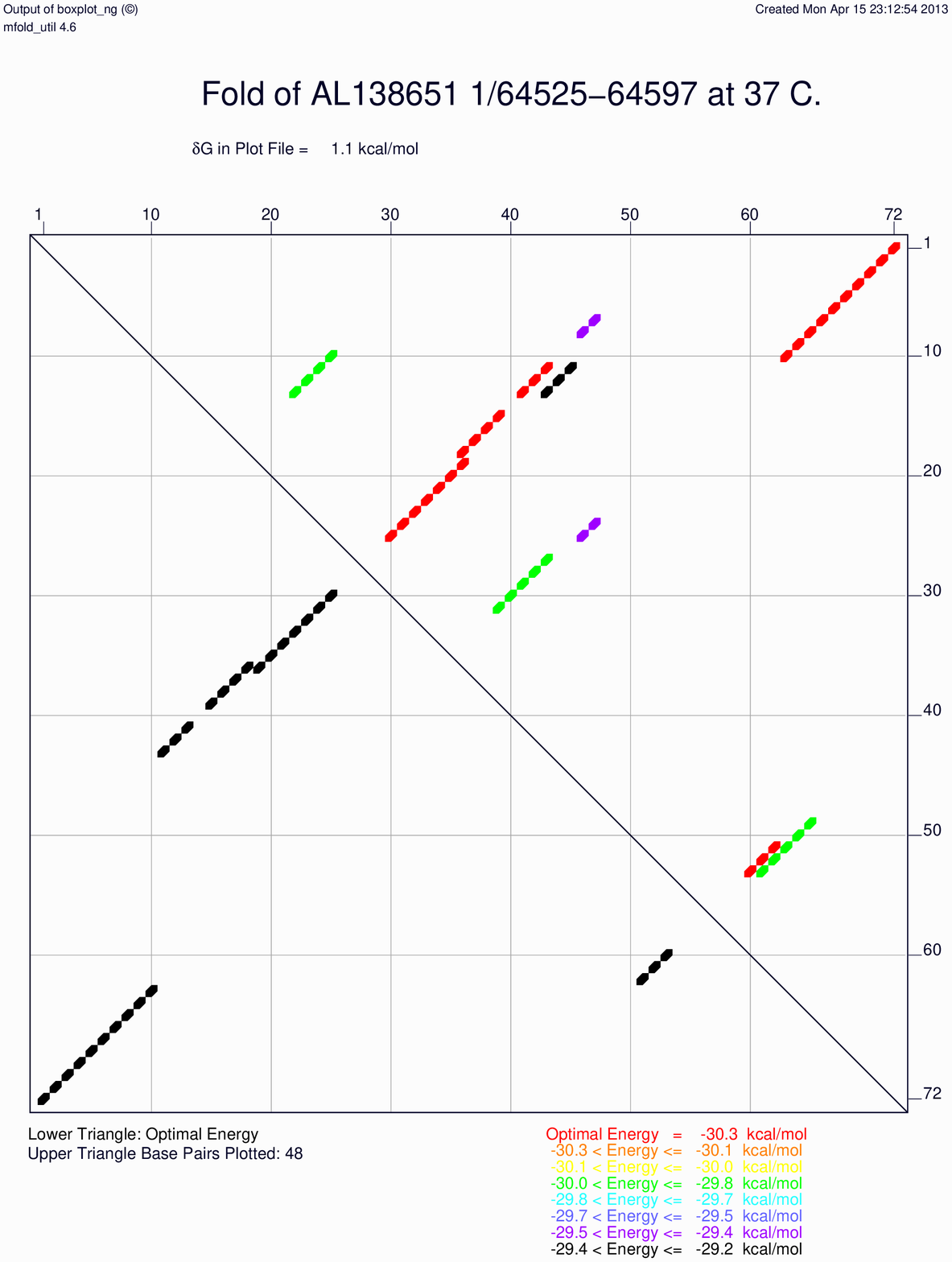}
    }
    \caption{\label{fig:energy_dot_plot}\small
      (Upper right) An energy dot plot for a tRNA sequence (AL138651.1/64525--64597), 
      produced by Mfold Web Server (http://mfold.rna.albany.edu/?q=mfold) \cite{pmid12824337}.
      Base-pairs with warmer colors have lower free energy, compared to the ones with  colder colors.
      (Lower left) The secondary structure of the MFE structure.
    }
  \end{figure}

  {
    \section{Evaluation measures in RNA secondary structure predictions}\label{sec:acc_meas_rna}%
  }
  The evaluation of a predicted RNA secondary structure is usually 
  based on {\em base-pairs}.
  First, we define 
  \begin{align*}
    &TP=\sum_{i<j}I(y_{ij}=1)I(\theta_{ij}=1),\ 
    TN=\sum_{i<j}I(y_{ij}=0)I(\theta_{ij}=0)\\
    &FP=\sum_{i<j}I(y_{ij}=1)I(\theta_{ij}=0),\ 
    FN=\sum_{i<j}I(y_{ij}=0)I(\theta_{ij}=1)
  \end{align*}
  for a reference secondary structure $\theta$ and
  a predicted structure $y$; $I(\cdot)$ is
  the indicator function, which takes a value of 1 or 0 depending 
  on whether the condition constituting its argument is true or false.
  These four parameters count the numbers of true positives (TP), 
  true negatives (TN), false positives (FP) and false negatives (FN). From these we can construct three accuracy measures that can be used for evaluation: the sensitivity (SEN), positive predictive value (PPV) and the Matthew Correlation Coefficient (MCC).
  \begin{align}
    &\mbox{SEN}=\frac{\mbox{TP}}{\mbox{TP}+\mbox{FN}}\label{eq:SEN}\\
    &\mbox{PPV}=\frac{\mbox{TP}}{\mbox{TP}+\mbox{FP}}\label{eq:PPV}\\
    &\mbox{MCC}=
    {\small \frac{\mbox{TP}\cdot \mbox{TN}-\mbox{FP}\cdot \mbox{FN}}
      {\sqrt{(\mbox{TP}+\mbox{FP})(\mbox{TP}+\mbox{FN})(\mbox{TN}+\mbox{FP})(\mbox{TN}+\mbox{FN})}}}\label{eq:MCC}.
  \end{align}
  There is a tradeoff between SEN and PPV, 
  while MCC is an accuracy measure which balances those two measures\footnote{F-score is sometimes 
    utilized instead of MCC, which is a balanced measure between Sensitivity and PPV.}.

  {
    \section{Generalized estimators of Section~\ref{sec:rna_sec_2}}\label{sec:max_pseudo_acc}
  }
  The approximate estimator {described in Section~\ref{sec:rna_sec_2}}
  can be adapted to utilize more general accuracy measures as follows.
  \begin{align}
    \hat y =\mathop{\mathrm{arg\ max}}_{y\in \mathcal{S}(x)} \sum_{\theta\in\mathcal{S}(x)}Acc(\theta,y)p(\theta|x)
  \end{align}
  where $Acc$ is a function of TP, TN, FP and FN (including MCC and F-score):
  \begin{align*}
    Acc = f(\mbox{TP},\mbox{TN},\mbox{FP},\mbox{FN}).
  \end{align*}
  For this general accuracy measure,
  {let} us define the {\em pseudo}-expected accuracy by
  \begin{align}
    \widehat{Acc}^0(y)=f(\widehat{\mbox{TP}},\widehat{\mbox{TN}},\widehat{\mbox{FP}},\widehat{\mbox{FN}}).\label{eq:pacc}
  \end{align}
  Finally, the estimator
  \begin{align}
    \hat y = \mathop{\mathrm{argmax}}_{y\in\mathcal{S}(x)} \widehat{Acc}^0(y).\label{eq:approx_acc}
  \end{align}
  is obtained, and the computation is conducted in combination 
  with stochastic sampling (Section~\ref{sec:ss}).

  \section{Supplementary Figure}

  %
  \begin{figure}[th]
    \centerline{
      \includegraphics[width=0.6\textwidth]{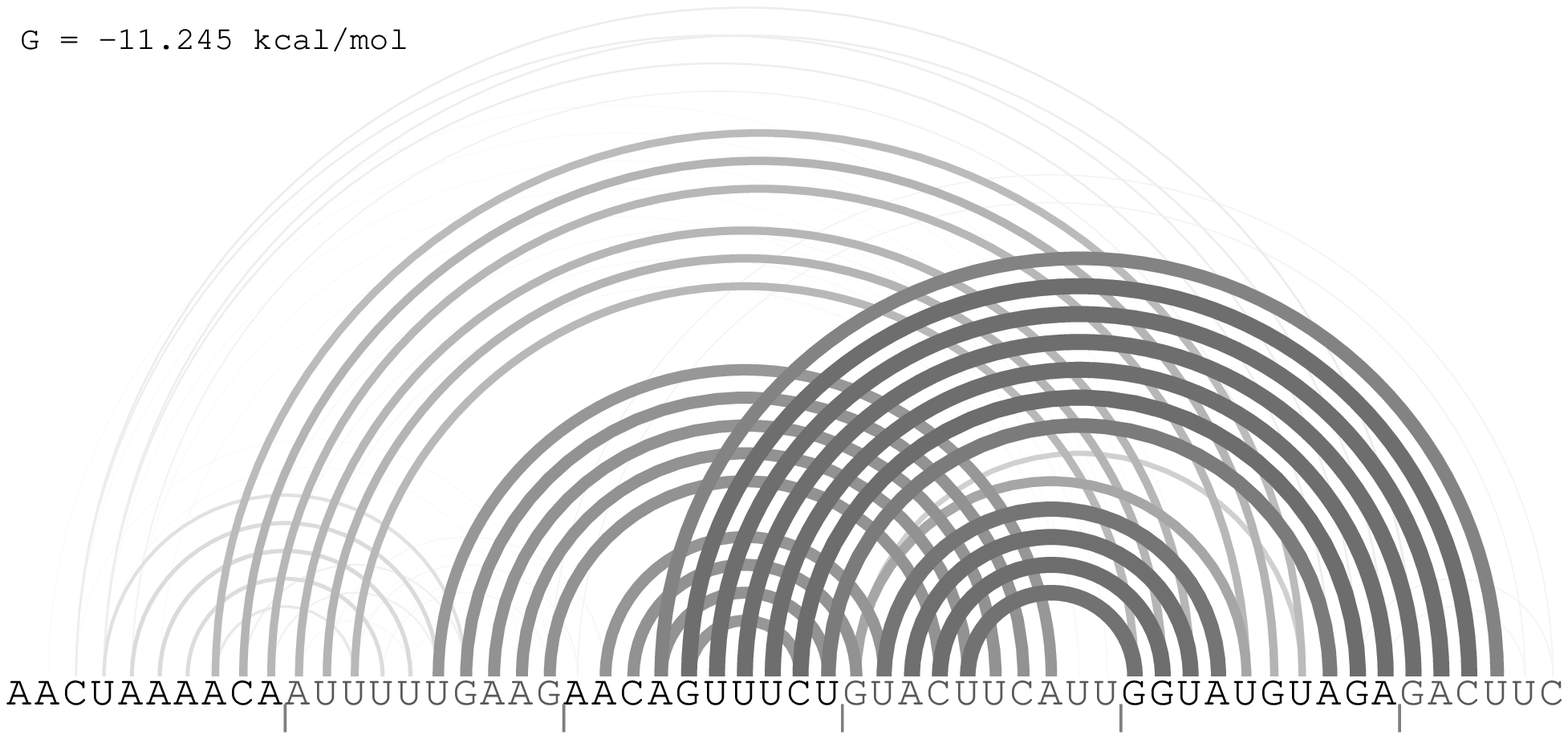}
    }
    \caption{\label{fig:rnabow}\small
      Visualization of RNA base-pairing probabilities by RNABow \cite{pmid23407410}.
    }
  \end{figure}

\end{document}